\documentclass[conference]{IEEEtran}
\usepackage{graphicx, times, amsmath, amsfonts, amsthm, comment}
\usepackage{amssymb,epstopdf}
\usepackage{algorithmic}
\usepackage{algorithm}
\usepackage{color, soul}
\usepackage{mathtools}

\newcommand{\beq}{\begin{equation}}
\newcommand{\eeq}{\end{equation}}
\newcommand{\tbf}{\textbf}
\newcommand{\tit}{\textit}

\newcommand{\ud}{\mathrm{d}}

\newcommand*{\mathcolor}{}
\def\mathcolor#1#{\mathcoloraux{#1}}
\newcommand*{\mathcoloraux}[3]{%
  \protect\leavevmode
  \begingroup
    \color#1{#2}#3%
  \endgroup
}

\theoremstyle{plain}
\newtheorem{propcounter}{Proposition}
\newtheorem{proposition}[propcounter]{Proposition}
\theoremstyle{plain}
\newtheorem{corocounter}{Corollary}
\newtheorem{corollary}[corocounter]{Corollary}
\theoremstyle{plain}

\theoremstyle{plain}
\newtheorem{assumecounter}{Assumption}
\newtheorem{assumption}[assumecounter]{Assumption}

\newcommand {\Fcal}{\mathcal{F}}

\newcommand {\Kcal}{\mathcal{K}}

\newcommand {\Ncal}{\mathcal{N}}

\begin{document}

\title{Inter-Operator Infrastructure Sharing: Trade-offs and Market}

\author{
\IEEEauthorblockN{Tachporn Sanguanpuak\IEEEauthorrefmark{1}, Sudarshan Guruacharya\IEEEauthorrefmark{2}, Ekram Hossain\IEEEauthorrefmark{2}, Matti Latva-aho\IEEEauthorrefmark{1}}
\IEEEauthorblockA{\IEEEauthorrefmark{1}Dept. of Commun. Eng., Univ. of Oulu, Finland; \IEEEauthorrefmark{2}Dept. Elec. \& Comp. Eng., Univ. of Manitoba, Canada.}
\IEEEauthorblockA{Email: \{tsanguan, matla\}@ee.oulu.fi; \{Sudarshan.Guruacharya, Ekram.Hossain\}@umanitoba.ca}
}\maketitle

\begin{abstract}

We model the problem of infrastructure sharing among mobile network operators (MNOs) as a multiple-seller single-buyer market where the MNOs
are able to share their own base stations (BSs) with each other. First, we use techniques from stochastic geometry to find the coverage probability of the infrastructure sharing system and analyze the trade-off between increasing the transmit power of a BS and the BS intensity of a buyer MNO required to achieve a given quality-of-service (QoS) in terms of the coverage probability. We also analyze the power consumption of the network per unit area (i.e., areal power consumption) and show that it is a piecewise continuous function composed of a linear and a convex functions. We show that when the transmit power of the BSs and/or the BS intensity of a network increases, the system becomes interference limited and the coverage probability tends to saturate at a certain value. As such, when the required QoS is set above this bound, an MNO can improve its coverage by buying infrastructure from other MNOs. Subsequently,  we analyze the strategy of a buyer MNO on choosing how many MNOs and which MNOs to buy the infrastructure from. The optimal strategy of the buyer is given by greedy fractional knapsack algorithm. On the sellers' side, the pricing and the fraction of infrastructure to be sold are formulated using a Cournot oligopoly game. 

\end{abstract}


\section{Introduction} \label{section:introduction}

In recent years, the concept of network infrastructure sharing has been investigated to address two kinds of concerns. On the one hand, with the growing demand for mobile services, the under utilization of dedicated spectrum auctioned off to the mobile network operators (MNOs) has become a bottleneck for the future growth of the industry \cite{Cisco2014}. While on the other hand, in areas or time periods where demand can be low, such as in rural areas or developing countries, or during night time, the high cost of network infrastructure forces the operators to charge high prices from their customers, making the mobile services unaffordable to most people, hence further driving down the demand \cite{ITU2016,GSMA2012}. 

One possible paradigm to address these issues is to allow the MNOs to share their infrastructures in order to maximize the use of existing network resources while simultaneously minimizing the cost and resources \cite{ITU2016,Matinmikko2014}. It also allows for a faster deployment of network services. Such sharing of infrastructure can be passive or active: Passive sharing refers to the sharing of physical space, such as buildings, sites, and masts. In active sharing, active elements of the network such as antennas, spectrum, entire base stations, or even elements of core network are shared. Thus, such active sharing allows mobile roaming, which allows an MNO to make use of another network in a place where it has no coverage or infrastructure of its own. 

There has been a growing number of work dedicated to investigate this issue. In \cite{Jorswieck2014}, hardware demonstration of the benefit of inter-operator spectrum sharing was demonstrated. In \cite{Kibilda2013}, infrastructure sharing was studied with full and partial coverage provisioning. A real-world multi-operator mobile network with infrastructure sharing was also shown to reduce significantly the number of base stations required to provide mobile service and improve coverage. In \cite{Kibilda2015}, stochastic geometry was used to investigate infrastructure sharing, spectrum sharing, and the combination of two. When both types of sharing is allowed, the authors showed that a trade-off existed between coverage and data rate performance. 
In \cite{Wang2016}, the authors also exploited stochastic geometry to study the trade-off involved in spectrum sharing and infrastructure sharing.

We consider multiple co-located deployment of network infrastructures by different MNOs, where the MNOs are assumed to operate over orthogonal frequency bands. In the infrastructure sharing deployment, each base station (BS) can be utilized by the users subscribed to more than one MNO. The MNO that installs the BS is considered as a potential seller of the BS infrastructure. This is the incumbent MNO. The entrant MNOs that use the BS of the incumbent MNO to serve its mobile user equipment (UE) are considered as the buyers. In the presence of multiple seller MNOs, it is assumed that they compete with each other to sell their infrastructure to a potential buyer. Our study only focuses on the sharing of infrastructure among the MNOs, that is, the MNOs do not share their spectrum. 

In this paper, we consider the scenario where there are multiple seller MNOs and one buyer MNO. In this case, we study the strategy of a buyer MNO, that decides which MNOs to buy the infrastructure from, and how much infrastructure to buy from them.
We propose a cost minimization problem for the buyer MNO, while guaranteeing the quality-of-service (QoS) to its UEs, in terms of the signal-to-interference-plus-noise ratio (SINR) coverage probability, as a fractional knapsack problem. Since there is a single buyer in the market, the competition among the buyers is not considered here.  

Next, we consider the market from the point of view of the sellers which compete with each other to sell the infrastructure. We model the competition among the seller MNOs as a Cournot-Nash game. The seller MNOs compete with each other in terms of their supply (a fraction of infrastructure to be shared), the cost associated (e.g., due to power consumption at the BSs), and the selling price with the objective to gain the highest profit.  As such we find the Cournot-Nash equilibrium and obtain the equilibrium price. We use results from stochastic geometric analysis of large-scale cellular networks to evaluate SINR outage probability and power consumption to model such market.

The major contributions of the paper are as follows:

\begin{itemize}
\item The paper presents an infrastructure sharing model with multiple seller MNOs and a single buyer MNO. The downlink SINR coverage probability, which is considered to be the QoS metric for the buyer MNO, is analyzed using stochastic geometry. Subsequently, the tradeoff between the transmission power of a BS and the BS deployment density for the buyer MNO  is analyzed. Also, for a seller MNO, since its profit depends on its cost of network operation, the areal power consumption (i.e., power consumption per unit area) at the BSs is analyzed. 
\item The optimal strategy for the buyer MNO, in order to minimize the cost of purchase, is obtained by solving a fractional knapsack problem. 
\item The optimal strategy for the seller MNOs, in terms of the fraction of  infrastructure to be shared and the pricing for the infrastructure, is obtained as the equilibrium of a Cournot-Nash game.
\end{itemize}

The rest of the paper is organized as follows. Section \ref{section:systemmodel} describes the system model. Section \ref{sec:stogeoana} gives the  stochastic geometrical analysis of the SINR outage probability of a typical user based on which the trade-off between BS transmit power and BS deployment is analyzed. Section \ref{section:strategybuyer} analyzes the strategic behavior of a buyer MNO to buy infrastructure from multiple seller MNOs. Section \ref{section:cournotseller} analyzes the competition among multiple sellers  using a Cournot-Nash game. The numerical results are presented in Section \ref{section:Numerical Results} before the paper is concluded in Section \ref{section:Conclusion}.

\section{System Model and Assumptions} \label{section:systemmodel}
Consider a system with $K+1$ MNOs given by the set $\Kcal = \{0, 1, \ldots, K\}$  to serve a common geographical area. We consider a single-buyer multiple-seller market for infrastructure sharing. Let MNO-$0$ denote our buyer MNO. Let the set of BSs owned by MNO-$k$ be given by $\Fcal_k$, where $k\in\Kcal$. Each of the BSs and user equipments (UEs) is assumed to be equipped with a single antenna. The maximum transmit power of each BS is $p_{\max}$. Also, a UE subscribing to an MNO  associates to the nearest BS belonging to that MNO. The BSs owned by different MNOs are spatially distributed according to homogeneous Poisson point processes (PPPs). Let the spatial intensity of BSs per unit area of MNO-$k$ be denoted by $\lambda_k$, where $k\in\Kcal$. Furthermore, each MNO-$k$, $k\in\Kcal$, is assumed to operate on orthogonal spectrum. Thus, there is no inter-operator interference among the MNOs.

During the sharing of infrastructure,  we assume the following statements to hold:

\begin{assumption}
When the buyer MNO-$0$ is allowed to use the infrastructure of a seller MNO-$k$, where $k\in \Kcal\backslash\{0\}$, the UEs of MNO-$0$ associates with the nearest available BSs owned by MNO-$0$ or the seller MNO-$k$.
\end{assumption}

\begin{assumption}
The buyer MNO-$0$ is assumed to use the infrastructure, but \tbf{not} the spectrum, belonging to a seller MNO-$k$, where $k\in \Kcal\backslash\{0\}$. As such a UE of MNO-$0$ served by the shared BS of a seller MNO-$k$ has to operate on the spectrum belonging to the MNO-$0$ itself. Since the seller MNO operates on a different spectrum, the shared BSs of the seller do not add extra interference to the UEs of MNO-$0$.\footnote{We assume that the MNO-0's users activity in seller MNOs is low and can be neglected.}
\end{assumption}

If the buyer MNO-$0$ shares infrastructure with $\Ncal \subseteq \Kcal\backslash\{0\}$ seller MNOs, then a UE subscribing to MNO-$0$ can effectively associate to any one of the enlarged set of BSs given by $\Fcal = \Fcal_0 \cup (\cup_{i \in \Ncal} \Fcal_i)$. This implies that the net BS intensity that a typical UE of MNO-$0$ will find itself in is
\beq
\lambda = \lambda_0 + \sum_{i \in \Ncal} \lambda_i,
\eeq
due to the superposition property of PPP. Note that despite the sharing of BSs among MNOs, there is no inter-operator interference among MNOs in our system model, since each MNO operates over a separate spectrum. Due to \textbf{Assumption 2}, the buyer will purchase only the infrastructure of the seller MNO and not the spectrum.


\section{SINR Coverage and Tradeoff Between Transmission Power and BS Deployment}
\label{sec:stogeoana}
\subsection{SINR Coverage Probability}
\label{subsec:avgcov}
Without loss of generality, we will consider a typical UE of MNO-$0$ to be located at the origin, which associates with the nearest BS in the enlarged set of BSs given by $\Fcal$. We will denote the nearest BS from $\Fcal$ to the typical UE as BS-$0$.  We assume that the message signal undergoes Rayleigh fading with the channel gain given by $g_0$. Furthermore, let $\alpha > 2$ denote the path-loss exponent for the path-loss model $r_0^{-\alpha}$, where $r_0$ is the distance between the typical UE and BS-$0$. Finally, let $\sigma^2$ denote the noise variance and $p$ denote the transmit power of all the BSs in MNO-$0$, including BS-$0$. The downlink ${\rm SINR}$ at the typical UE is
\beq
{\rm SINR} = \frac{g_0 r_0^{-\alpha}p}{I + \sigma^2}.
\eeq
Here $I = \sum_{i\in\Fcal_0 \backslash \{0\}} g_i r_i^{-\alpha}p$ is the interference experienced by a typical UE only from the BSs that operate on the spectrum of MNO-$0$. These are the BSs  that belong only to MNO-$0$ itself. Here $g_i$ is the co-channel gain between typical UE and interfering BS-$i$ and $r_i$ is  the distance between the typical UE and the interfering BS-$i$, where  $i\in\Fcal_0 \backslash \{0\} $.  The transmit power of each BS is $0 < p \leq p_{\max}$.

Given a detection threshold $T$, if ${\rm SINR} < T$ the UE is said to experience an outage. Likewise, if ${\rm SINR} > T$, then the UE is said to have coverage. The SINR coverage probability for a typical UE of MNO-$0$'s cellular network is defined as
\beq 
P_c = \mathrm{Pr}({\rm SINR} > T).
\eeq

In \cite[Theorem 1]{Andrews2011}, the authors derived a formula for the coverage probability of a typical UE when the BS are distributed according to a homogeneous PPP of intensity $\lambda$  as given by
\begin{equation}
P_c = \pi \lambda \int_0^{\infty}\exp \{-(Az + Bz^{\alpha /2})\} dz,
\label{eqn:coverage-integral}
\end{equation}
where $A = \pi \lambda \beta$ and $B = \frac{T \sigma^2}{p}$, and
\[ \beta = \frac{2 (T/p)^{2/\alpha}}{\alpha} \mathbb{E}_{g}[g^{2/\alpha} (\Gamma(-2/\alpha, T g/p))- \Gamma(-2/ \alpha)].  \]
When the interfering links undergo Rayleigh fading, $\beta = 1 + \rho(T,\alpha)$, where 
\beq
 \rho(T,\alpha) = T^{2/\alpha} \int_{T^{-2/\alpha}}^\infty (1+u^{\alpha/2})^{-1} \ud u.
 \label{eqn:rho} 
 \eeq
For this special case, we see that $\beta$ is independent of transmit power.

However, for our system, due to the fact that the interference does not scale with the BS intensity, we have to modify the above formula. We can proceed in a manner similar to the proof of \cite[Theorem 1]{Andrews2011} and show that a more general  coverage formula is given as follows:

\begin{proposition}
The coverage probability of a typical UE of buyer MNO-$0$ under the Assumptions 1 and 2 is
\begin{equation}
P_c = \pi \lambda \int_{0}^{\infty} \exp \{- (A' z + B z^{\alpha/2}) \} dz,
\label{eqn:modified-coverage-integral}
\end{equation}
where $A' = \pi (\lambda - \lambda_0 (1 - \beta))$ and $\lambda = \lambda_0 + \sum_{i\in\Ncal} \lambda_i$ such that $\Ncal \subseteq \Kcal \backslash \{0\}$. Here $B$ and $\beta$ are the same as in (\ref{eqn:coverage-integral}).
\end{proposition}
\begin{IEEEproof}
Proceed in a manner similar to the proof of \cite[Theorem 1]{Andrews2011}, keeping in mind that interference is contributed only from BSs of MNO-$0$, while BS association is contributed by all MNOs in $\Ncal$.
\end{IEEEproof}


\begin{corollary}
When there is no infrastructure sharing,  $\Ncal = \emptyset$, (\ref{eqn:modified-coverage-integral}) reduces to (\ref{eqn:coverage-integral}). 
\label{coro:mod-to-old-cov-integral}
\end{corollary}

Except for $\alpha = 4$, the integral for $P_c$ cannot be evaluated in closed form. Nevertheless, a simple closed-form approximation for the general case, where $\alpha > 2$, and where both noise and intra-operator interference are present, can be given as \cite[Eqn. 4]{Sudarshan2016}
\begin{align}
P_c \simeq \pi \lambda \left[ A'+ \frac{\alpha}{2} \frac{B^{2/\alpha}}{\Gamma\big(\frac{2}{\alpha}\big)} \right]^{-1}.
\label{eqn:coverage-approx}
\end{align}

\begin{proposition}
Let $N = |\Ncal|$. Then, (i) for fixed $N$, $\lim_{\lambda_0 \rightarrow \infty} P_c = 1/ \beta$, (ii) for fixed $\lambda_0$, if $\lim_{N\rightarrow \infty} \sum_{i\in\Ncal} \lambda_i = \infty$, then  $\lim_{N\rightarrow \infty}P_c = 1$, (iii) for fixed $N$, if $\lambda_0 = 0$, then $P_c \simeq \left[1 + \frac{\alpha}{2\pi \Gamma(\frac{2}{\alpha})} \frac{B^{2/\alpha}}{\sum_{i\in\Ncal} \lambda_i}\right]^{-1}$. 
\label{prop:limits_of_Pc}
\end{proposition}
\begin{IEEEproof}
(i) From the approximation in (\ref{eqn:coverage-approx}), we see that as $\lambda_0 \rightarrow \infty$, since $B$ and $\sum_{i=0}^N \lambda_k$ remains constant,  $P_c \rightarrow 1/\beta$. (ii) Again from the approximation (\ref{eqn:coverage-approx}), since $B$ and $\lambda_0(1-\beta)$ are constants, $P_c \rightarrow 1$. (iii) When $\lambda_0 = 0$, $A' = \pi \sum_{i\in\Ncal} \lambda_i$. Simplifying (\ref{eqn:coverage-approx}), we obtain  the desired result.
\end{IEEEproof}

In our case, the increase in BS intensity does not correspond with the increase in co-channel interference, which is different from \cite{Andrews2011}. \tbf{Proposition \ref{prop:limits_of_Pc}} also confirms our intuition that greater sharing of infrastructure leads to better coverage. However, paradoxically, increasing one's own infrastructure leads to degrading of performance.

\subsection{Minimum Transmit Power Required to Satisfy the QoS}
\label{subsec:Tx_infra}
Let us further assume that the MNO-$0$ wants to ensure that the coverage probability of a typical UE satisfies the QoS constraint
\beq
P_c \geq 1 - \epsilon,
\label{eqn:QoS-constrain}
\eeq
where $0< \epsilon < 1$ is some arbitrary value. 

In order to satisfy the coverage constrain in (\ref{eqn:QoS-constrain}), the minimum power required for each BS of MNO-$0$, for given infrastructure, is given by the following proposition.

\begin{proposition}
Assume that the interfering links undergo Rayleigh fading and $\lambda$ be defined as before. Then, given that $1- \epsilon < 1/\beta'$, where $\beta' = 1 - \lambda_0(1-\beta)/\lambda$, the minimum transmit power required for each BS of MNO-$0$ such that $P_c \geq 1 - \epsilon$, is 
\beq 
p \simeq c \lambda^{-\alpha/2},  
\label{eqn:min-power}
\eeq 
where $c = \left[\frac{2 \pi ( 1- (1-\epsilon)\beta' ) }{\alpha (1-\epsilon)( T \sigma^2)^{2/\alpha}} \Gamma(\frac{2}{\alpha}) \right]^{-\alpha / 2}$.
\label{prop:power-infra-trade-off}
\end{proposition}  

\begin{IEEEproof}
When the interfering links undergo Rayleigh fading, $\beta = 1 + \rho$, as given in (\ref{eqn:rho}), and is independent of $p$. Thus, using (\ref{eqn:coverage-approx}) in the inequality $P_c \geq 1 - \epsilon$, and solving for $p$, we obtain the desired result. For $p > 0$, it suffices that $1 - (1-\epsilon)\beta' > 0$ in the expression for $c$. Re-arranging the terms gives the sufficient condition.
\end{IEEEproof} 
 
In Proposition \ref{prop:power-infra-trade-off}, note that $1/\beta' \geq 1/\beta$.

\subsection{Trade-off Between Power and Infrastructure}
\label{subsec:useroutage}
Every MNO wishes to guarantee a certain probability of coverage to its own customers. For this purpose, if a UE is experiencing outage, the MNO can either choose to increase the transmit power of the BSs so as to increase the coverage radius, or offload the call to a shared BS. It is natural to wonder at the possible trade-off between increasing the power and sharing more infrastructure. The answer was provided by \tbf{Proposition \ref{prop:power-infra-trade-off}}. 

Intuitively, in \textbf{Proposition~\ref{prop:power-infra-trade-off}}, the minimum required transmit power decreases with increasing BS densification. As for the sufficiency condition, $1/\beta'$ is the maximum attainable coverage probability as transmit power $p \rightarrow \infty$ and as the system becomes interference limited, $B \rightarrow 0$. That is, we have the upper bound $P_c \leq 1/\beta'$. Thus, the QoS, $1 - \epsilon$, can be achieved by varying the transmit power only when $1- \epsilon < 1/\beta'$. If this condition is violated, the QoS cannot be satisfied by simply varying the transmit power, and MNO-$0$ will have to buy more infrastructure from other MNOs. A few special cases are worth mentioning: (1) For fixed $N$, if $\lambda_0 \rightarrow \infty$,  then $\beta' \rightarrow \beta$. (2) For fixed $\lambda_0$, if $\lambda \rightarrow \infty$, then $\beta' \rightarrow 1$. 

Let $R$ be the cell radius of a BS defined as the distance at which a UE will receive $-3$ dB SNR. Then, for the important special case when there is no infrastructure sharing,  we have the following scaling law as a corollary.

\begin{corollary}
When there is no infrastructure sharing, the minimum BS transmit power for which $P_c \geq 1 - \epsilon$ is $p \simeq c \lambda_0^{-\alpha/2},$ where $c$ is independent of $\lambda_0$. Also, the optimal cell radius is $ R \simeq \frac{c'}{\sqrt{\lambda}},$ where $c' = (2c/\sigma^2)^{1/\alpha}$.
\label{coro:power-infra-trade-off}
\end{corollary}
\begin{IEEEproof}
From Proposition \ref{prop:power-infra-trade-off}, since $\Ncal = \emptyset$, we have $\lambda = \lambda_0$ and $\beta' = \beta$. Also, since the cell edge is defined as the distance at which SNR is $-3$ dB, we have $\frac{p R^{-\alpha}}{\sigma^2} = \frac{1}{2}$. Putting $p = c \lambda_0^{-\alpha/2}$, we can solve for $R$ to obtain the result.
\end{IEEEproof}

A scaling law similar to \textbf{Corollary \ref{coro:power-infra-trade-off}} can be found in \cite[Lemma 1]{Sarkar2014} and \cite[Lemma 1]{Perabathini2014} for homogeneous PPP, using a slightly different approximation, as $p \propto \lambda_0^{-\alpha/2+1}$. However, our formula differs from theirs in the order of the exponent as well as the proportionality constant.  Likewise, the scaling law for the  cell radius, $R \propto 1/\sqrt{\lambda_0}$, corresponds to that obtained by \cite{Hanly2002} for hexagonal grid model.

\subsection{Areal Power Consumption}
Let the transmit power of each BS belonging to the seller MNO-$k$, where $k \in \Kcal \backslash \{0\}$, be denoted by $p_k$. Apart from the transmit power, each BS also consumes a fixed amount of circuit power, denoted by $p_c$. Hence, the total power consumption of a  BS of an MNO-$k$ is $p_k + p_c$. Since the MNO-$k$ has $\lambda_k$ BS per unit area, the areal power consumption of the network (i.e., power consumption per unit area) is  
\beq
S_k = \lambda_k ( p_k + p_c).
\label{eqn:power-per-area}
\eeq  

For MNO-$k$, let the QoS constaint on coverage probability of a typical UE be $P_c \geq 1 - \epsilon$ and the threshold SINR be $T_k$. In order to satisfy this constrain, it can either increase its BS intensity $\lambda_k$ or increase its transmit power $p_k$. The trade-off between $\lambda_k$ and $p_k$ was given by \textbf{Proposition \ref{prop:power-infra-trade-off}}. Similarly, the trade-off between $\lambda_k$ and $S_k$ follow immediately.

\begin{proposition}
Given the assumptions in Proposition \ref{prop:power-infra-trade-off}, areal power consumption of seller MNO-$k$, where $k \in \Kcal \backslash \{0\}$, is 
\beq
S_k(\lambda_k) = \left\{ \begin{array}{ll}
\lambda_k(p_{\max} + p_c), & \mathrm{if} \quad  0 \leq \lambda_k \leq (\frac{c_k}{p_{\max}})^{2/\alpha}, \\
\lambda_k(c_k \lambda_k^{-\alpha/2} + p_c), & \mathrm{if} \quad  \lambda_k \geq (\frac{c_k}{p_{\max}})^{2/\alpha},
\end{array} \right.
\label{eqn:areal-power}
\eeq
where $c_k = \left[\frac{2 \pi ( 1- (1-\epsilon)\beta ) }{\alpha (1-\epsilon)( T_k \sigma^2)^{2/\alpha}} \Gamma(\frac{2}{\alpha}) \right]^{-\alpha / 2}$. 
\end{proposition}
\begin{IEEEproof}
Since the MNO-$k$ does not buy infrastructure from other MNOs, the net BS intensity that a typical UE of MNO-$k$ experiences is $\lambda_k$. Thus, from Corollary \ref{coro:power-infra-trade-off}, we have $p_k \simeq c_k \lambda_k^{-\alpha/2}$. Putting $p_k$ in (\ref{eqn:power-per-area}) and recalling that $0 < p_k \leq p_{\max}$, we have (\ref{eqn:areal-power}).
\end{IEEEproof}

 We see that $S_k$ is a piece-wise continuous function of $\lambda_k$. The $S_k$ initially increases linearly with $\lambda_k$. Beyond a certain point, $S_k$ behaves as a convex function. This can be verified by checking the second derivative of $S_k$ for $\lambda_k \geq (\frac{c_k}{p_{\max}})^{2/\alpha}$ as 
\[ \frac{\ud^2 S_k}{\ud \lambda_k^2} = \frac{c_k \alpha (\alpha - 2)}{4} \lambda_k^{-\frac{\alpha}{2} - 1}. \]
Since $c_k >0$ and $\alpha > 2$, we have $\frac{\ud^2 S_k}{\ud \lambda_k^2} > 0$, proving the convexity of $S_k$ in the region $\lambda_k \geq (\frac{c_k}{p_{\max}})^{2/\alpha}$. As such, studying the behaviour of $S_k$ is not straightforward.  Nevertheless, the local minima in the convex region can be found.

\begin{proposition}
Given the assumptions in Proposition \ref{prop:power-infra-trade-off}, let $\lambda_{th} = (\frac{c_k}{P_{\max}})^{2/\alpha}$. Then, for the region $\lambda_k \geq \lambda_{th}$, the BS intensity for which the areal power consumption of MNO-$k$, where $k \in \Kcal \backslash \{0\}$, is minimum is 
\beq
\lambda_{k,\min} = \max\left(\lambda_{th},\left[\frac{c_k}{p_c}\left(\frac{\alpha}{2} -1 \right)\right]^{2/\alpha}\right).
\eeq
\end{proposition}
\begin{IEEEproof}
We have $\ud S_k/\ud \lambda_k = p_c - (c_k(\alpha - 2) \lambda_k^{-\alpha/2})/2$. Solving $\ud S_k/\ud \lambda_k = 0$ for $\lambda_k$, we have $\lambda_k^{*} = [\frac{c_k}{p_c}(\frac{\alpha}{2} -1)]^{2/\alpha}$. This is clearly the minima if $\lambda_{th} < \lambda_k^{*}$. Otherwise, $\lambda_{k,\min} = \lambda_{th}$. 
\end{IEEEproof}

\begin{figure}[h]
\centering
\includegraphics[height=3.1 in, width=3.1 in, keepaspectratio = true]{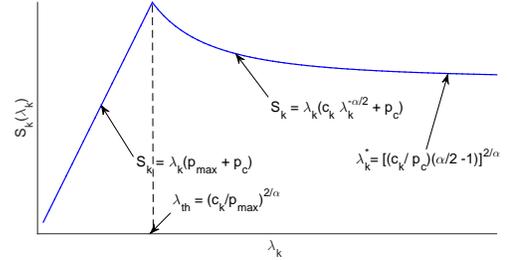}
\caption{The areal power consumption ($S_k$) versus BS intensity ($\lambda_k$).}
\label{fig:Sk_vs_lambdak}
\end{figure}

In Fig. \ref{fig:Sk_vs_lambdak}, we illustrate $S_k$ as a function of $\lambda_k$, as given in (\ref{eqn:areal-power}). We can see that it is composed of linear and convex parts. The convex part of $S_k$ corresponds to that obtained for hexagonal grid models via simulations in \cite{Richter2009}. Similar, but not the same, formulas have been given in \cite{Sarkar2014,Perabathini2014}.

\section{Buyer's Strategy}\label{section:strategybuyer}
In this section, we propose a strategy for the buyer MNO-$0$ which will allow it to choose the seller MNOs to buy the infrastructure from. By using our method, the buyer MNO will select the minimum number of seller MNOs, at minimum cost, such that it can serve its UEs guaranteeing some QoS. We first have the following proposition:

\begin{proposition}
For the QoS condition $P_c \geq 1 - \epsilon$ to be feasible for the buyer MNO-$0$, the net BS intensity $\lambda = \lambda_0 + \sum_{i \in \Ncal} \lambda_i$ must satisfy 
\beq
\lambda \geq \frac{1 - \epsilon}{\epsilon}(\gamma - \lambda_0(1 - \beta)),
\label{eqn:delta}
\eeq
where $\gamma = \frac{\alpha}{2\pi} \frac{B^{2/\alpha}}{\Gamma\big(\frac{2}{\alpha}\big)}$, for some $\Ncal$ such that $\emptyset \subseteq \Ncal \subseteq \Kcal$.
\end{proposition}
\begin{IEEEproof}
Using the approximation (\ref{eqn:coverage-approx}) in (\ref{eqn:QoS-constrain}) and solving for $\lambda$, we have required result.
\end{IEEEproof}


If there is no cost attached to the infrastructure sharing, or if the cost of buying infrastructure from all the seller MNOs is the same, then the QoS constrain (\ref{eqn:QoS-constrain}) can be easily satisfied by selecting the $N$ MNOs with largest BS intensities $\lambda_i$ such that (\ref{eqn:delta}) is satisfied. This greedy approach thus gives the required MNOs from whom to buy the infrastructure from. However, if there is a cost associated with the sharing of infrastructure, then we can formulate a cost minimization problem with the QoS constraint, which can be written as a linear program as follows:
\begin{align}
\label{eqn:knapsack}
\text{min} & \sum_{k \in \Kcal \backslash \{0\} } q_{k} x_{k} \\
\text{s.t.} \quad \mathrm{(C1)} & \quad \sum_{k \in \Kcal \backslash \{0\} } \lambda_{k} x_{k} \geq  \frac{1- \epsilon}{\epsilon} \left( \gamma - \lambda_0 (1-\beta) \right) - \lambda_0, \nonumber 
\end{align}
where $q_{k}$ is the price of infrastructure when buying from MNO-$k$, where $k \in \Kcal \backslash \{0\}$, and $x_k$ ($0 \leq x_{k} \leq 1$) denotes the fraction of infrastructure bought from seller MNO-$k$. We can interpret $x_k$ in two possible ways: 1) The buyer MNO-$0$ buys the entire infrastructure of MNO-$k$ but utilizes the whole infrastructure of MNO-$k$ for only $x_k$ fraction of time, 2) the MNO-$0$ buys only a fraction $x_k$ of the total infrastructure of MNO-$k$, but utilizes it all the time. 

The problem in (\ref{eqn:knapsack}) is an instance of a \tit{knapsack problem}. 
In the knapsack interpretation of problem (\ref{eqn:knapsack}), the seller MNOs are interpreted as ``items'', their BS intensities are interpreted as ``weights'', and the right hand term of constrain (C1) is interpreted as ``weight capacity'' of a ``bag''. Likewise, $q_k$ is interpreted as the ``value'' of the $k$-th ``item''. Since $x_k \in [0,1]$, the problem $(\ref{eqn:knapsack})$ becomes a \tit{fractional knapsack problem}, and a greedy algorithm can be used to obtain the optimal solution \cite[Chap 17.1]{Korte2012}. 

The greedy algorithm is provided in \textbf{ Algorithm \ref{alg:FracKnapsack}}. The idea behind this greedy algorithm is as follows: We first sort the seller MNOs according to their cost per BS intensity in ascending order. We then select an MNO in that order if its weight (the BS intensity) is less than or equal to the residual weight capacity of the knapsack. In our case, the maximum weight capacity of knapsack is defined by $\bar{w} = \frac{1- \epsilon}{\epsilon} \left(\gamma - \lambda_0 (1-\beta) \right) - \lambda_0$. If the BS intensity exceeds $\bar{w}$, then the buyer MNO-$0$ buys only a fraction of the infrastructure from the remaining seller MNOs. Hence, we can define the variable $x_k$ as
\beq
x_k = \left\{\begin{array}{cr}
				1, & \mbox{if} \quad \lambda_i \leq \bar{w}-w  \\
				\frac{(\bar{w}-w)}{\lambda_i}, & \mbox{if} \quad \lambda_i > \bar{w}-w,
			 \end{array} \right.
\eeq
where $w$ is the weight in the knapsack thus far.
\begin{algorithm}
\caption{Fractional Knapsack Algorithm}
\label{alg:FracKnapsack}
 \begin{algorithmic}[1]
 \STATE Initialize $x_k =0$, $w = 0$, and $V = 0$
 \STATE Compute $\rho_k= q_k/\lambda_k$
  \STATE Sort the sellers by $\rho_k$ in ascending order such that $\rho_{\pi_1} \leq \rho_{\pi_2} \cdots \leq \rho_{\pi_K}$
    \FOR{$i = 1$ \TO $K$}
        \IF{$\lambda_{\pi_i} \leq \bar{w} - w$}
            \STATE $x_{\pi_i} = 1$
             \STATE $V = V + q_{\pi_i}$
		\STATE $w = w + \lambda_{\pi_i}$
         \ELSE
            \STATE $x_{\pi_i} = \frac{\bar{w}-w}{\lambda_{\pi_i}}$
            \STATE $V = V + q_{\pi_i} y_{\pi_i}$
		\STATE Terminate
      \ENDIF
 \ENDFOR
 \end{algorithmic}
 \end{algorithm}
%

\section{Sellers' Competition: Pricing of Infrastructure} \label{section:cournotseller}
In this part, we will study the equilibrium pricing due to the sellers competition as well as the optimal fraction of infrastructure that the seller MNOs will be willing to sell. We will formulate the seller competition as a Cournot-Nash oligopoly game \cite{Friedman1983}.

Let the fraction of infrastructure to be sold from the seller MNO-$k$, $k\in \Kcal \backslash \{0\}$, be $z_k$, where $0 \leq z_k \leq 1$. Then, the total amount of infrastructure sold by the seller MNO-$k$ is $y_k = \lambda_k z_k$. Let the cost of operating its infrastructure be $C_k(y_k)$, which we can define as
\beq 
C_k(y_k) = a_k S_k(y_k) + d_k,
\label{eqn:Ck}
\eeq 
where $a_k$ is the price of areal power consumption, $d_k$ is a fixed operation cost, and $S_k$ is as given in (\ref{eqn:areal-power}).

Let the overall infrastructure from $K$ seller MNOs available in the market be denoted by $y = \sum_{k=1}^K y_k $. Also, let us denote the fraction of infrastructure of all MNOs except MNO-$k$ by $y_{-k} = y - y_{k}$. Let the selling price of the infrastructure be $Q(y)$. The price function $Q(y)$ is assumed to be monotonically increasing with the total supply of infrastructure, in accordance to the ``law of supply". We will assume $Q(y)$ to be 
\beq
Q(y) = \theta + \eta y,
\label{eqn:revenueQ}
\eeq  
where $\theta$ is the initial installation price of infrastructure from all seller MNOs and $\eta$ denotes the marginal price of the total  infrastructure $y$ in the market.  
Thus, the MNO-$k$'s profit is 
\begin{align}
F_{k}(y_1,\ldots,y_k) = y_k Q(y) - C_k( y_k).
\label{eqn:profit}
\end{align}

In order to maximize the profit of MNO-$k$ with respect to $y_k$, we first partial differentiate (\ref{eqn:profit}) with respect to $y_k$, and noting that $\partial y/ \partial y_k = 1$, we obtain
\beq
\frac{\partial F_k}{\partial y_k}  = y_k \frac{\ud Q}{\ud y} + Q - \frac{\ud C_k}{\ud y_k}.
\label{eqn:partial-deri}
\eeq

Using the optimality condition $\frac{\partial F_k}{\partial y_k} = 0$ in (\ref{eqn:partial-deri}) and solving for $y_k$, we obtain 
\beq
y_k = \frac{1}{\frac{\ud Q}{\ud y}} \left(\frac{\ud C_k}{\ud y_k} - Q \right),
\label{eqn:BR}
\eeq
which is in a fixed-point form. Let us denote the function at the right hand side of (\ref{eqn:BR}) by $\mathrm{BR}_k(y_{-k}) \equiv \frac{1}{\frac{\ud Q}{\ud y}} \left(\frac{\ud C_k}{\ud y_k} - Q \right)$, which we referred to as the best response of MNO-$k$ to the action of other competitive sellers. 

Here we have $\frac{\ud Q}{\ud y} = \eta$, and 
\[ \frac{\ud C_k}{\ud y_k} = \left\{\begin{array}{ll}
				a_k( p_{\max} + p_c),    & \mathrm{if} \quad 0 \leq y_k \leq (\frac{c_k}{p_{\max}})^{2/\alpha}  \\
				a_k(1 - \frac{\alpha}{2}) c_k y_k^{-\alpha/2} + a_k p_c,  & \mathrm{if} \quad  y_k \geq (\frac{c_k}{p_{\max}})^{2/\alpha}.
			 \end{array} \right.\]
We see that the marginal cost of MNO-$k$ is constant up until a certain point, after which the marginal cost starts to monotonically increase. Thus, the action of MNO-$k$ to sell $y_k$ amount of infrastructure depends on the action of other MNOs, as given by the equation $y_k = \mathrm{BR}_k(y_{-k})$. Substituting $\frac{\ud C_k}{\ud y_k}$, $\frac{\ud Q}{\ud y}$ and $Q$ in (\ref{eqn:BR}), and recalling that $y = y_k +y_{-k}$, we obtain the best response of MNO-$k$ as
\beq y_k  = \left\{\begin{array}{ll}
				\frac{U_k}{2} - \frac{y_{-k}}{2}, & \mathrm{if} \quad 0 \leq y_k \leq (\frac{c_k}{p_{\max}})^{2/\alpha}  \\
				\frac{V_k y_k^{-\alpha/2}}{2} + \frac{W_k}{2} - \frac{y_{-k}}{2},  & \mathrm{if} \quad y_k \geq (\frac{c_k}{p_{\max}})^{2/\alpha},
			 \end{array} \right.
\label{eqn:BRk}
\eeq
where $U_k = \frac{a_k (p_{\max} + p_c)-\theta}{\eta}$, $V_k = \frac{a_k (1-\frac{\alpha}{2})c_k}{\eta}$ and $W_k = \frac{a_k p_c - \theta}{\eta}$.   

The equilibrium solution of the Cournot-Nash oligopoly market, $\mathbf{y}^*$, is the fixed point of the best response function. As such, the best responses of all $K$ seller MNOs can be expressed in vector form as $\mathbf{y}^{*} = \mathrm{BR}(\mathbf{y}^{*})$, where $\mathbf{y^{*}} = [y_1^{*}, y_2^{*}, \ldots, y_K^{*}]^{T}$ and $\mathrm{BR}(\tbf{y}^{*}) = [\mathrm{BR}_1(y_{-1}^{*}), \mathrm{BR}_2(y_{-2}^{*}), \ldots, \mathrm{BR}_K(y_{-K}^{*})]^{T}$. The $[.]^{T}$ denotes transpose of vector.  By taking summation of (\ref{eqn:BRk}) over all $K$ seller MNOs, and using the fact that $\sum_{k \in \Kcal \backslash \{0\}} y_{-k} = \sum_{k \in \Kcal \backslash \{0\}}(y - y_k) = (K-1)y$, we can analytically solve the equilibrium quantity $y^{*}$ as
\beq
 y^{*}  = \left\{\begin{array}{ll}
				\frac{\sum_{k\in \Kcal \backslash \{0\}}U_k}{K+1},   & 0 \leq y_k^{*} \leq (\frac{c_k}{p_{\max}})^{\frac{2}{\alpha}}  \\
				\frac{\sum_{k \in \Kcal \backslash \{0\}} ( V_k (y_k^{*})^{-\alpha/2} + W_k)}{K+1} , & y_k^{*} \geq (\frac{c_k}{p_{\max}})^{\frac{2}{\alpha}}.
			 \end{array} \right.
\label{eqn:Equi_quan}
\eeq    
Once the equilibrium quantity $y^*$ is computed, we can find the corresponding equilibrium price $q^*$ by substituting $y^*$ into the price function in (\ref{eqn:revenueQ}), we get $q^* = Q(y^*)$.

\section{Numerical Results}\label{section:Numerical Results}
We assume that the BSs are spatially distributed according to homogeneous PPP inside a circular area of $500$ meter radius for all $K+1$ MNOs. The seller MNOs are assumed to have the same intensity of BSs per unit area. The maximum transmit power of each BS is $p_{\max} = 10$ dBm, the SINR threshold at each user is $T_k = 15$ dB, the path-loss exponent is $\alpha = 4$, and noise $\sigma^2 = -120$ dBm.

\subsection{Effect of Changing the Outage QoS}

\begin{figure}[h]
\centering
\includegraphics[height=3.3 in, width=3.4 in, keepaspectratio = true]{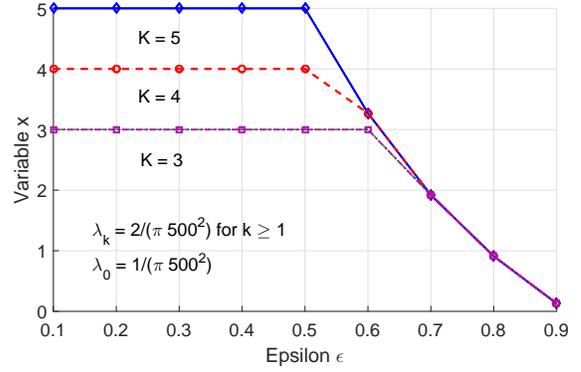}
\caption{The fractional $x$ for $\sum_{i=1}^{K}x_i$ versus the tolerable outage probability.}
\label{fig:Fractional_epsilon}
\end{figure}

In Fig. \ref{fig:Fractional_epsilon}, we illustrate the fraction of infrastructure bought by MNO-$0$ while increasing the values of tolerable outage probability $\epsilon$ (i.e. $P_c \geq 1 - \epsilon$). Each BS from all MNOs are assumed to transmit at its the maximum power. We also assume that the price of infrastructure $q_k$ is the same for all sellers. The fractional variable $x = \sum_{i=1}^{K} x_i$ indicates the proportion of infrastructure that MNO-$0$ has bought. We see that for low values of  $\epsilon$, MNO-$0$ cannot satisfy the required QoS solely through own infrastructure. In this figure, the MNO-$0$ needs to buy infrastructure from all the sellers. When $\epsilon$ increases beyond a certain value, $x$ starts to decrease, indicating that at higher $\epsilon$ MNO-$0$ buys less infrastructure.

\subsection{Effect of Changing the BS Intensity of MNO-$0$}

\begin{figure}[h]
\centering
\includegraphics[height=3.3 in, width=3.4 in, keepaspectratio = true]{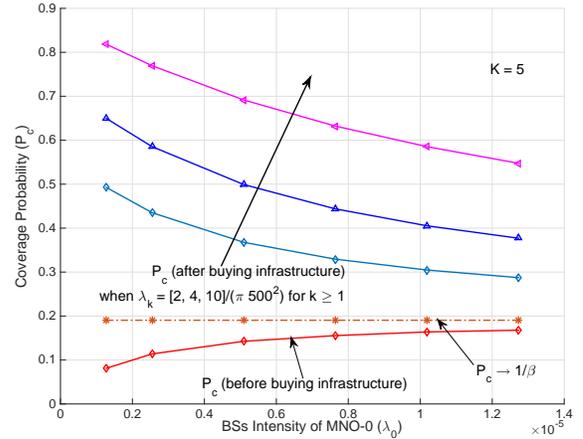}
\caption{The coverage probability of user of MNO-$0$ before and after buying infrastructure.}
\label{fig:Pc}
\end{figure}

In Fig. \ref{fig:Pc}, the difference between coverage probability of the buyer MNO-$0$ before and after buying infrastructure is shown. Each BS of every MNOs transmits at maximum power. We assume that the number of seller MNOs is $K=5$ and that all of them have same BSs intensity. Before buying infrastructure, it can be seen that the $P_c$ of MNO-$0$ approaches $1/\beta$ as $\lambda_0$ increases. The MNO-$0$ cannot simply increase its own BS intensity to achieve a coverage more than the upper bound $1/\beta$, as we proved in \textbf{Proposition~2~(i)}. The MNO-$0$ will have to buy more infrastructure to gain more coverage. After buying infrastructure from all five MNOs, we see that the coverage of MNO-$0$ improves and is greater than $1/\beta$. For fixed $\lambda_0$, when the BS intensity of  seller MNO-$k$, $k\in\Kcal\backslash \{0\}$, increases, the coverage of MNO-$0$ also increases. This verifies \textbf{Proposition~2~(ii)}. Also, for fixed $\lambda_k$, where $k \geq 1$, as $\lambda_0$ increases, the coverage of MNO-$0$ decreases, in accordance to \textbf{Proposition~2~(i)}.

%

\subsection{The Market Equilibrium Price and Quantity}
%

\begin{figure}[h]
\centering
\includegraphics[height=3.3 in, width=3.4 in, keepaspectratio = true]{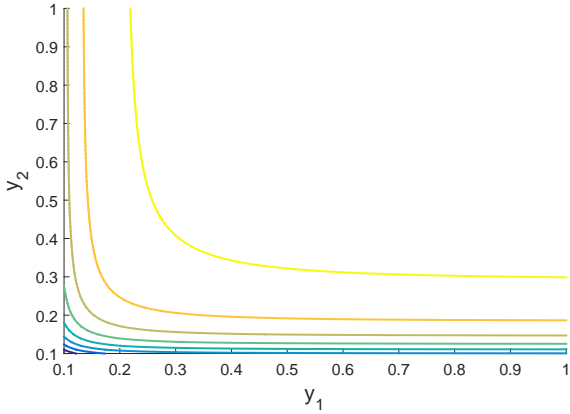}
\caption{Equilibrium quantity ($y^{*}$) with $y_1$ and $y_2$.}
\label{fig:EquiQuantity_y1y2}
\end{figure}

\begin{figure}[h]
\centering
\includegraphics[height=3.3 in, width=3.4 in, keepaspectratio = true]{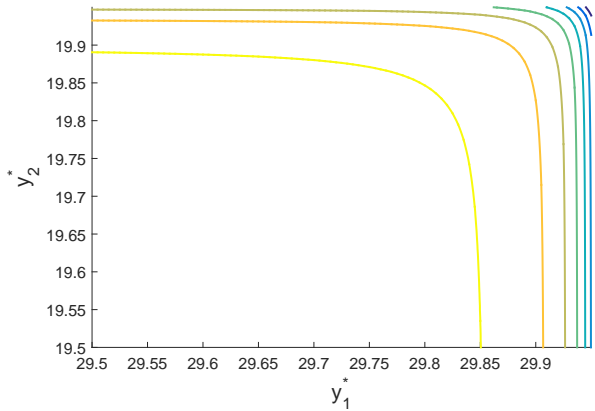}
\caption{Equilibrium Price ($q^{*}$) with $y_1^{*}$ and $y_2^{*}$.}
\label{fig:EquiPrice_BR}
\end{figure}

The market equilibrium quantity and price are illustrated in Fig.~\ref{fig:EquiQuantity_y1y2} and Fig.~\ref{fig:EquiPrice_BR}. We consider two seller MNOs, where the best response of seller MNO-$1$ to the action of the MNO-$2$ and vice versa can be obtained from (\ref{eqn:BRk}). In Fig.~ \ref{fig:EquiQuantity_y1y2}, we have plotted the equilibrium quantity $y^{*}$ from (\ref{eqn:Equi_quan}) with respect to $y_1^*$ and $y_2^*$. We observe that for fixed $y^*$, the equilibrium contour lines for large $y_1^*$ and $y_2^*$ are hyperbolas. Fig.~\ref{fig:EquiPrice_BR} shows the equilibrium price with respect to $y_1^*$ and $y_2^*$. The contour lines of equilibrium price gives a single equilibrium price solutions.


\section{Conclusion}\label{section:Conclusion}
We studied the infrastructure trading problem for multiple seller MNOs and one buyer MNO using stochastic geometry. We first analyzed the coverage probability of the buyer MNO, and studied the trade-offs between buying of  infrastructure and increasing of transmit power. We then focused on the strategy of buyer and the competition between sellers. We provided the strategy of a buyer MNO on choosing how many MNOs and which MNOs to buy infrastructure from in order to satisfy the QoS. The strategy of the buyer was formulated as a fractional knapsack problem and the optimal solution was found via greedy algorithm. The pricing and fraction of infrastructure that sellers are willing to sell is formulated using Cournot-Nash oligopoly game. One major conclusion is: infrastructure sharing can improve cellular coverage.


This work can be extended in several directions: 1)  in addition to infrastructure sharing, spectrum sharing can be also considered, 2)  multiple buyer MNOs can be considered where they compete with each other to obtain their demand with lowest price, in addition to the seller competition. 

\bibliographystyle{IEEE}

\end{document}